\documentstyle[12pt]{article}
\def\as{\alpha_S}
\def\ee{e^+e^-}
\def\b0{\beta_0}

\def\frac#1#2{ {{#1} \over {#2} }}
\def\rat#1#2{\mbox{\small $\frac{#1}{#2}$}}
\def\VEV#1{\left\langle #1\right\rangle}
\def\beq{\begin{equation}}
\def\beeq{\begin{eqnarray}}
\def\eeq{\end{equation}}
\def\eeeq{\end{eqnarray}}

\def\np#1#2#3{Nucl.\ Phys.\ B#1 (19#3) #2}

\def\prl#1#2#3{Phys.\ Rev.\ Lett.\ #1 (19#3) #2}

\begin{document}
\begin{titlepage}
\renewcommand{\thefootnote}{\fnsymbol{footnote}}
\begin{flushright}
     Cavendish--HEP--94/7 \\
     hep-ph/9408222
\end{flushright}
\vspace*{\fill}
\begin{center}
{\Large \bf
Estimation of Power Corrections\\[1ex]
to Hadronic Event Shapes\footnote{Research supported in part by
the U.K. Particle Physics and Astronomy Research Council and by
the EC Programme ``Human Capital and Mobility", Network ``Physics at
High Energy Colliders", contract CHRX-CT93-0537 (DG 12 COMA).}}
\end{center}
\par \vskip 2mm
\begin{center}
        {\bf B.R.\ Webber} \\
        Cavendish Laboratory, University of Cambridge,\\
        Madingley Road, Cambridge CB3 0HE, U.K.
\end{center}
\par \vskip 2mm
\begin{center} {\large \bf Abstract} \end{center}
\begin{quote}
Power corrections to hadronic event shapes are estimated using a
recently suggested relationship between perturbative and non-perturbative
effects in QCD.  The infrared cutoff dependence of perturbative calculations
is related to non-perturbative contributions with the same dependence on
the energy scale $Q$. Corrections proportional to $1/Q$ are predicted,
in agreement with experiment. An empirical proportionality between the
magnitudes of perturbative and non-perturbative coefficients is noted.
\end{quote}
\vspace*{\fill}
\begin{flushleft}
     Cavendish--HEP--94/7\\
     July 1994
\end{flushleft}
\end{titlepage}
\renewcommand{\thefootnote}{\fnsymbol{footnote}}

\par \vskip 15mm \noindent
Event shapes in the process $\ee\to$ hadrons have been widely used to test
QCD and to determine its coupling constant $\as$. Predictions of infrared safe
quantities from perturbation theory, either in next-to-leading order or with
enhanced terms summed to all orders, generally provide a good description of
the data, provided they are subjected to ``hadronization corrections" obtained
from non-perturbative models.  For most quantities these corrections appear
empirically to be proportional to $1/Q$ where $Q$ is the centre-of-mass
energy.  This is in contrast to the total cross section, which for massless
quarks has a leading power correction of order $1/Q^4$.  The
smallness of non-perturbative effects in the total cross section and
related quantities, such as the hadronic widths of the $Z^0$ boson and
the $\tau$ lepton, has led to a preference for these quantities as a
means of determining $\as$, even though event shapes have a stronger
perturbative dependence on $\as$.

In the case of the total cross section, we also have some understanding
of the leading power correction. It is believed to arise from the vacuum
expectation value of the gluon condensate, $\VEV{\as G^2}$, which is
the relevant gauge-invariant operator of lowest dimension, giving a
correction proportional to  $\VEV{\as G^2}/Q^4$.  For event
shapes, we do not even know why the corrections should be of order
$1/Q$: there are no operators of dimension one to which they could
be related.

Another way of discussing power corrections is in terms of
renormalons [\ref{MW}]. These are singularities of the Borel
transform of the all-orders perturbative expression for a quantity,
generated by a factorial growth of the perturbation series at high
orders.\footnote{For a recent review, see Ref.~[\ref{mueller}]}
Such growth in the soft region is thought to give rise
to an infrared renormalon at the position $8\pi/\b0$ ($\b0=11-\rat 2 3 N_f$)
in the Borel plane, corresponding to a power correction proportional to
\beq
\exp\left[-\frac{8\pi}{\b0\as(Q^2)}\right]
\sim \frac{\Lambda^4}{Q^4}
\eeq
where $\Lambda$ represents the
QCD scale. The existence of the renormalon is supposed to
indicate that the full QCD prediction would exhibit the same
power correction. In this language, the appearance of $1/Q$ corrections
to event shapes would be associated with a new infrared renormalon at
$2\pi/\b0$ in the Borel transforms of these quantities.

In the present paper, I apply an idea due to Bigi, Shifman, Uraltsev
and Vainshtein [\ref{BSUV}], who argue that there is a simple
correspondence between renormalon positions and the power corrections
to fixed-order perturbative predictions evaluated with an infrared cutoff.
In $\ee\to$ hadrons, to first order in $\as$ all diagrams are QED-like and
a suitable cutoff can be imposed by introducing a small mass $\mu$
in the denominator of the gluon propagator.\footnote{Recall that
in QED a photon mass can be introduced in this way without violating
the Ward identities associated with current conservation.}
One finds that the first-order perturbative total cross section
with such a cutoff, normalized to the  Born value, is
\beq\label{Rp}
R_p = 1+\as/\pi - D\,\as\,\mu^4/Q^4 + \ldots
\eeq
where $D$ is a constant and the ellipsis represents non-leading power
corrections.  The dependence on $\mu$ must cancel between
this expression and the soft contribution, which builds the renormalon.
Thus the leading renormalon occurs at $8\pi/\b0$, as described above,
and we expect a non-perturbative contribution of the form
\beq
R_{np} = [C\,\Lambda^4 + D\,\as(\mu)\,\mu^4]/Q^4
\eeq
where $C$ is a constant. The $\mu$-dependence appears as an arbitrariness
in the part of the correction that we attribute to the renormalon and
the part that is generated in fixed order.

In the case of event shapes, however, one finds that the introduction
of a gluon mass in the way outlined above leads to corrections of order
$\as\mu/Q$. That is, for a generic (infrared safe) event shape $S$
we find in first order
\beq\label{Sp}
S_p = A_S\,\as - D_S\,\as\,\mu/Q + \ldots
\eeq
instead of a relation of the form (\ref{Rp}).
The coefficients $D_S$ are easily computed; to this order, they arise
entirely from the reduction of phase space for real gluon emission.
Their values, together with those of the leading coefficients $A_S$,
are listed for some representative quantities
in Table~\ref{hadro}. Here $T$ is the thrust [\ref{thrust}],
$C$ is the $C$-parameter [\ref{cpar}], and $\sigma_L$ is the longitudinal
cross section [\ref{sigl}].

\begin{table}
\renewcommand{\arraystretch}{1.5}
\begin{center}
\begin{tabular}{|c|c|c|c|} \hline
      $S$   & $A_S$   & $D_S$ & $C_S\,\Lambda$ \\ \hline\hline
$\VEV{1-T}$ & $0.335$ & $\rat{16}{3\pi} = 1.7$  & $\sim 1.0$ GeV
\\ \hline
$\VEV{C}$   & $1.375$ & 8     & $\sim 5.0$ GeV
\\ \hline
$\sigma_L$  & $1/\pi$ &$4/3$ & $\sim 0.8$ GeV
\\ \hline
\end{tabular}
\end{center}
\caption{\label{hadro}
 Coefficients of terms in Eqs.~(\ref{Sp}) and (\ref{Snp}). }
\end{table}

{}From Eq.~(\ref{Sp}) we expect that a new infrared renormalon
at $2\pi/\b0$ is present in event shapes, leading to
a non-perturbative contribution
\beq\label{Snp}
S_{np} = [C_S\,\Lambda + D_S\,\as(\mu)\,\mu]/Q \; ,
\eeq
whose dependence on the arbitrary cutoff $\mu$ cancels
against that of the perturbative part, leaving
a cutoff-independent power correction $C_S\,\Lambda/Q$.
The observed value of this correction, inferred [\ref{glasgow}]
from experimental data, is shown for each quantity in Table~1.

It is remarkable that the observed $1/Q$ corrections are,
within the uncertainties, proportional to the perturbative coefficients
$D_S$, suggesting that Eq.~(\ref{Snp}) takes the general form
\beq\label{Snp2}
S_{np} = C_S\frac{\Lambda}{Q}\,\left[1
           + d\,\as(\mu) \frac{\mu}{\Lambda}\right]
\eeq
where $d$ is a constant, roughly equal to 0.5 if we take $\Lambda\simeq 0.3$
GeV.  Since $\as(\mu)\,\mu/\Lambda$ has a minimum value (at $\mu=e\Lambda$)
of $2\pi e/\b0\simeq 1.9$, the cutoff-dependent contribution is
comparable to, or greater than, the full $1/Q$ correction.
This makes the applicability of the method marginal: it would
be preferable if there existed a region of $\mu$ in which the cutoff
dependence was small compared with both the perturbative and
non-perturbative contributions, as discussed in Ref.~[\ref{BSUV}].

It would obviously be of interest to apply the above approach to
a wide variety of quantities, and to try to construct event
shapes from which $1/Q$ corrections are absent. From Table~1
we see that the combination $\VEV{T+2C/3\pi}$ might be of
this type. It would be desirable to extend the treatment
to higher orders in perturbation theory, but it is difficult
to see how this can be done in a gauge-invariant way, unless the
dimensional regularization method can be adapted to the purpose.

I am grateful to Steven Bass and Mike Seymour for comments.

\par \vskip 15mm
\noindent{\large \bf References}
\begin{enumerate}
\item  \label{MW}
       A.V.\ Manohar and M.B.\ Wise, Univ.\ of California at San Diego
       preprint UCSD/PTH 94-11.
\item  \label{mueller}
       A.H.\ Mueller, in {\em QCD 20 Years Later}, vol.~1 (World Scientific,
       Singapore, 1993).
\item\label{BSUV}
       I.I.\ Bigi, M.A.\ Shifman, N.G.\ Uraltsev and A.I.\ Vainshtein,
       Univ.\ of Minnesota preprint TPI-MINN-94/4-T (CERN-TH.7171/94).
\item\label{thrust}
       E.\ Farhi, \prl{39}{1587}{77}.
\item\label{cpar}
       R.K.\ Ellis, D.A.\ Ross and A.E.\ Terrano, \np{178}{421}{81}.
\item\label{sigl}
       P.\ Nason and B.R.\ Webber, \np{421}{473}{94}.
\item\label{glasgow}
       B.R.\ Webber, to appear in {\em Proc.\ 27th Int.\ Conf.\ on High
       Energy Physics}, Glasgow, 1994; see also {\em Proc.\ Workshop on
       New Techniques for Calculating Higher-Order QCD Corrections}
       (ETH, Zurich, 1992).
\end{enumerate}
\end{document}